\documentstyle[12pt]{article}

\begin{document}
\setlength{\parskip}{5mm}

\title{A New Redshift Interpretation }
\author{Robert V. Gentry
      \\The Orion Foundation
      \\P.O. Box 12067
      \\Knoxville, TN 37912
       }
\date{astro-ph/9806280}

\maketitle

\begin{abstract}

A nonhomogeneous universe with vacuum energy, but without spacetime
expansion, is utilized together with gravitational and Doppler redshifts as
the basis for proposing a new interpretation of the Hubble relation and the
2.7K Cosmic Blackbody Radiation.

\end{abstract}

Hans-Dieter Radecke recently noted$^1$ that modern cosmology's dependence on
``interpretations of interpretations of observations'' means that ``We
should not to fall victim to cosmological hubris, but stay open for any
surprise.'' We now report what seems a major surprise---namely, the
discovery of a New Redshift Interpretation (NRI) of the Hubble redshift
relation and 2.7K Cosmic Blackbody Radiation (CBR) without assuming either
the Friedmann-Lemaitre wavelength expansion hypothesis or the Cosmological
Principle, the latter being long acknowledged as the ``...one great
uncertainty that hangs like a dark cloud over the standard model.''$^2$
Whereas the standard model and the NRI both interpret nearby galactic
redshifts as Doppler shifts, they differ significantly in their
interpretation of distant redshifts. This difference can be traced to two
fundamentally different views of the universe. The standard model utilizes a
universe governed by expanding-spacetime general relativity whereas the NRI
is based on a universe governed by static-spacetime general relativity. A
brief review of early twentieth-century astronomy and cosmology assists in
focusing more precisely on the nature of this difference.

In 1917 Einstein applied his newly developed static-spacetime general theory
of relativity to cosmology,$^3$ and introduced a cosmological constant to
maintain the universe in what was then thought to be a static condition. But
Edwin Hubble's momentous 1929 discovery$^4$ that galactic redshifts increase
in proportion to their distance challenged the static universe concept. His
discovery confronted cosmologists with two surprises, and they were
initially unprepared to deal with either. First, they were unaware of any
static-spacetime redshift interpretation which could account for increasing
galactic redshifts in a real, finite-density universe. Secondly, if Hubble's
results were interpreted as Doppler shifts they implied omnidirectional
galactic recession, which in turn implied the existence of a universal
center near the Galaxy.

In any event, whatever efforts cosmologists might have put forth to obtain a
static-spacetime interpretation of Hubble's discovery were effectively cut
short when their attention was soon directed to the potential cosmological
implications of the hitherto virtually unnoticed results of Alexander
Friedmann$^5$ and Georges Lemaitre,$^6$ both of whom had found
expanding-spacetime solutions of the Einstein field equations in the early
and mid-1920s. Their results were attractive for two reasons. First, uniform
spacetime expansion showed promise for eliminating the implication of the
Galaxy occupying a preferred position in the universe. Hubble spoke for most
cosmologists of his time when he forthrightly admitted an extreme distaste
for such a possibility, saying it should be accepted only as a last resort.$%
^7$

Second, Lemaitre hypothesized that, apart from the well-known redshift due
relative motion of source and observer, expanding-spacetime should cause
photons everywhere to experience continuous, in-flight wavelength expansion
proportional to the expansion itself.$^6$ Thus was born the concept of
spacetime expansion redshifts, given by $z_{\exp }=\Re _o/\Re _e-1,$ where $%
\Re _o$ and $\Re _e$ represent the magnitudes of the postulated
Friedmann-Lemaitre spacetime expansion factors at observation and emission.$%
^6$

Despite its critical role in standard model theory, the foregoing expression
for $z_{\exp }$ is unique in that the physical existence of $\Re $ has never
been verified by experiment; the reason is that no method has yet been
proposed to measure $\Re ,$ either past or present$.$ Even so, expansion
redshifts have become the cornerstone of the standard model for two
reasons---namely, (1) because the experimentally determined Hubble redshift
relation, $z=Hr/c$, can be developed as a theoretical consequence of
spacetime expansion theory if the hypothesized expansion redshifts, $z_{\exp
}=\Re _o/\Re _e-1$, are assumed to be identical with $z_{obs}=\lambda
_o/\lambda _e-1,$ the observed redshifts of distant galaxies, and (2)
because of their key role in providing what has previously been thought to
be a unique interpretation of the 2.7K CBR. That interpretation assumes the
much earlier existence of a primeval fireball radiation wherein
matter/radiation decoupling occurred at about 3000K when the expansion
redshift was about 1000 compared to the present. It follows that a 1000-fold
redshifting of such a radiation by spacetime expansion would result in the
presently observed 2.7K CBR. The Hubble relation and 2.7K CBR scenarios are
widely understood as confirming the existence of expansion redshifts. Such a
conclusion may be premature, however, seeing that the crucially important
expansion factor, $\Re ,$ has yet to be experimentally verified.

This brings us to the standard model's second fundamental assumption known
as the Cosmological Principle---namely, that in the large scale the universe
is homogeneous and isotropic, or put in simpler terms, it is everywhere
alike. This Principle was earlier noted to be the ``...one great uncertainty
that hangs like a dark cloud over the standard model.''$^2$ Uncertainty
exists because, even though the Hubble relation is powerful evidence for
large-scale isotropy about the Galaxy, we simply cannot confirm universal
homogeneity because we lack knowing whether the Hubble relation would result
if redshift measurements were made from points of observation on other
galaxies.

Nevertheless the standard model requires homogeneity because in it galaxies
are assumed to be comoving bodies in expanding spacetime. That is, since
spacetime expansion is assumed to be uniform, comoving galactic separation
must likewise be uniform, which implies that all observers, regardless of
location, should see the same general picture of the universe. This is what
the standard model requires, and it is observationally unprovable.

In summary, then, our mini-review of twentieth century astronomy and
cosmology have revealed two reasons why we cannot be absolutely certain of
Friedmann-Lemaitre expansion redshifts and the standard model's cornerstone
postulate of a no-center universe governed by expanding-spacetime \smallskip
general relativity. First, the universal homogeneity required by the
standard model is acknowledged to be observationally unprovable. Second,
despite the fact that in theory all photons in the universe should be
synchronously experiencing in-flight wavelength expansion in direct
proportion to the instantaneous value of $\Re ,$ until now little attention
has been given to finding a method to test this prediction. More on this
later. For the present we say only that the foregoing uncertainties are
sufficient to suggest the possibility that the universe may not be governed
by expanding-spacetime general relativity required by the standard model. As
far as is known this paper is the first attempt to seriously explore the
cosmological consequences of such a possibility and, as will now be seen,
the results do appear quite surprising.

The foregoing account provides the basis for understanding why the NRI
attempts to account for the Hubble relation and the 2.7K CBR by using
Doppler and gravitational redshifts embedded in a universe governed by
static-spacetime general relativity. Without expanding spacetime there can
be no Cosmological Principle, and without this Principle the Hubble relation
implies the existence of a center in the NRI. In it the Hubble redshifts are
now interpreted solely in terms of relativistic Doppler and Einstein
gravitational redshifts, all cast within the framework of a finite,
nonhomogeneous, vacuum-gravity universe with cosmic center (C) near the
Galaxy.

The NRI framework assumes the widely dispersed galaxies of the visible
universe are enclosed by a thin, outer shell of hot hydrogen at a distance R
from the Galaxy. Thus, the volume of space enclosed by this luminous
shell---assumed, for ease of calculation, to have a nearly uniform
temperature of 5400K---would completely fill with blackbody cavity
radiation. But the radial variation of gravitational potential within this
volume means the cavity radiation temperature measured at any interior point
would depend on the magnitude of the Einstein gravitational redshift between
that point and the outer shell. By including relativistic vacuum energy
density, $\rho _v$, and pressure, p$_v$, into the gravitational structure of
the cosmos we now show how 5400K radiation emitted at R could be
gravitationally redshifted by a factor of 2000 so as to appear as 2.7K
blackbody cavity radiation here at the Galaxy.

In particular, if p$_v$ is negative, then, as Novikov$^8$ shows, $\rho _v$
will be positive, and the summed vacuum pressure/energy contributions to
vacuum gravity will be $-2\rho _v$. So, excluding the spherical hydrogen
shell at R, the net density throughout the cosmos from C to R would be $\rho
-2\rho _v$, where $\rho $ is the average mass/energy density. Beyond R both
densities are assumed to either cancel or exponentially diminish to
infinitesimal values, which effectively achieves for the NRI framework what
Birkhoff's theorem did for standard cosmology. This framework is sufficient
to compute the gravitational potentials needed to calculate both Hubble and
2.7K CBR redshifts in the NRI framework.

If $\Phi (0)$ and $\Phi (R$) represent the universal potentials at C and R,
then, $\Phi (R)$ $=$ $-(G/R)[M_1$ $+$ $M_S]$, and $\Phi (0)=-G[\int_0^R4\pi
(\rho -2\rho _v)r\cdot dr+M_S/R],$ ~where M$_S$ is the mass of the thin,
outer shell, and the net mass from C to R is $M_1=4\pi R^3(\rho -2\rho
_v)/3. $ To find M$_S$ we employ the boundary condition $\Phi (0)=0$, which
first yields $M_S=-2\pi R^3(\rho -2\rho _v)$ and then, by substitution, the
expression $\Phi (R)$ $=$ $2\pi GR^2[\rho -2\rho _v]/3$. Explanation of $%
\Phi (0)=0$ is given in ref. (9). Gravitationally redshifting the uniform
5400K radiation at R so as to appear as the 2.7K CBR at C means that $%
T_R=5400K$ and $T_C=2.7K$, in which case the gravitational redshift at C
would be,
\begin{equation}
z+1=\sqrt{1+2\Phi (0)/c^2}/\sqrt{1+2\Phi (R)/c^2}=T_R/T_C=5400/2.7.
\end{equation}
If, as soon to be explained, we let $\rho _v=8.790 \times 10^{-30}$ g/cm$^3$
and $\rho =2\times 10^{-31}$ g/cm$^3$, then Eq. (1) yields $R\cong \
1.362\times 10^{28}$ cm, or about $14.24 \times 10^9$ ly. These are the
conditions which allow the 2.7K CBR to be interpreted as blackbody cavity
radiation. Now consider the Hubble relation.

Interpreting the Hubble relation in the NRI framework assumes that distant
galaxies may have both radial and transverse velocity components relative to
C. Thus the appropriate redshift equation is given by the standard
expression,$\ ^{10}$
\begin{equation}
z=\ [1-\vec{v}\cdot \hat{k}/c]\sqrt{1+2\Phi (0)/c^2}/\sqrt{1+2\Phi (r)/c^2
-v^2/c^2} -1
\end{equation}
which describes how light emitted from a distant galaxy moving at velocity $%
\vec{v}$---with the unit vector $\hat{k}$ pointing from source to
observer---will be redshifted by a combination of gravitational and special
relativistic Doppler effects. In Eq. (2) $r$ is the distance from C to an
arbitrary galaxy where the gravitational potential is,
\begin{equation}
\Phi (r) =-G[(r^{-1}\int_0^r4\pi (\rho
-2\rho _v)r^2\cdot dr)+\int_r^R4\pi (\rho -2\rho _v)r\cdot dr+M_S/R]
\end{equation}

Substituting the expression $M_S$ $=-2\pi R^3(\rho -2\rho _v)$,
we arrive at $2\Phi (r)/c^2$ $=$ $4\pi Gr^2[\rho -2\rho _v]/3c^2$, which
together with the radial velocity, $v_r=-\vec{v}\cdot \hat{k}$, and boundary
condition $\Phi (0)=0,$ allows Eq. (2) to be rewritten as,
\begin{equation}
z=[1+v_r/c]/\sqrt{1+4\pi Gr^2(\rho -2\rho _v)/3c^2-v^2/c^2}-1.
\end{equation}
To obtain the Hubble relation from Eq. (4) we note that if $2\rho _v\ >
\rho $, then the $(\rho -2\rho _v)\ $density factor will cause any galaxy
located at a distance r from C to experience an outward radial acceleration, 
$\ddot{r}=-GM/r^2$, due to the enclosed negative mass $M=4\pi r^3(\rho
-2\rho _v)/3$. This leads to the equation $\ddot{r}=br$, where $b=4\pi
G(2\rho _v-\rho )/3$. Its solution is $r=r_g\exp \sqrt{b}t$, where $r_g$ is
a galaxy-specific, initial condition parameter. Taking its proper time
derivative leads to the expression $v_r=\dot{r}=\sqrt{b}r$, which is of
course the Hubble velocity-distance relation, $v_r=\ Hr$, with $H=\sqrt{b}.$

If $\rho =2\times 10^{-31}$ g-cm$^{-3}$ and $H=68$
km-s$^{-1}$-Mpc$^{-1}=2.203\times 10^{-18}$s$^{-1}$ are substituted into
$H^2=b=4\pi
G(2\rho _v-\rho )/3$, then the vacuum density is $\rho _v\cong 8.79\ \times
10^{-30}\ g$-$cm^{-3}$. This value for $H$ was earlier utilized to calculate
R $\cong $ 14.24 billion ly. And substitution of this value of the vacuum
density in $\rho _vc^2=Gm_{v\ }^6c^4/\hbar ^4$, as given by Novikov$^8$,
yields the average mass, m$_v$, of the virtual particles emitted by the
vacuum. If $m_e$ is the electron mass, then m$_v\ \simeq 82m_e$ for $\rho
=2\times 10^{-31}g$-$cm^{-3}$.

If we consider the case for dark matter, where $\rho =2\times 10^{-30}g$-$%
cm^{-3},$ then $\rho _v\cong 9.79\ \times 10^{-30}\ g$-$cm^{-3{ }}$and m%
$_v\ \simeq 83m_e.$ Thus the use of dark matter density in the NRI produces
only negligible changes in $m_v$ and $\rho _v$, which has the important
consequence of justifying the use of $\rho _{v{ }}$ as an adjustable
parameter in the expression $H^2=4\pi G(2\rho _v-\rho )/3$. Indeed, because
the right side of this expression occurs explicitly in the NRI redshift
Equation (4), it's possible to restate it in terms of $H^{2{ }}$ without
a density term. (See further discussion in the next paragraph.) This means
the results already obtained, as well as those that now follow, apply
equally to a universe with or without the assumption of dark matter.

Further interpretation of Eq. (4) requires evaluating the $\ v^2/c^2=$ $%
(v_r^2+v_\theta ^2)/c^2$ term. The most general case assumes a possible
transverse velocity component such that $v_\theta =u_gv_r$---where $u_g$ is
an undetermined, galactic-specific parameter constrained only by the $v<c$
condition--- thus leading to $\ v^2/c^2=(1+u_g^2)v_r^2/c^2=\
(1+u_g^2)(Hr/c)^2$. Thus, Eq. (4) can be rewritten as,
\begin{equation}
z=(1+Hr/c)/\sqrt{1-(2+u_g^2)(Hr/c)^2}-1.
\end{equation}

For small $r$, Eq. (5) reduces to the Hubble redshift relation, $z=Hr/c$,
which means the NRI\ framework successfully interprets both it and the 2.7K
CBR. Even the latter's microtemperature variations$^{11}$ can be interpreted
as temperature variations in the 5400K outer shell. Going further, the
equation, $\ddot{r}=\ br,$ with b as given above, essentially duplicates$%
^{12}$ the inflationary expansion relation, $\ddot{\Re}=\{8\pi G\rho
_f/3\}\Re $.

The NRI framework also succeeds in interpreting the observed variation of
CBR temperature with redshift. When Songaila et al.$^{13}$ investigated this
topic, their measurement yielded a CBR temperature of $7.4\pm 0.8$ K for $%
z=1.776.$ The NRI framework's prediction is obtained by substituting $r$ for 
$R$ in Eq. (1), from which it follows that the CBR temperature should vary
spatially with $r$ and $z$ as $T_z=T_C/\sqrt{1+2\Phi (r)/c^2}=(z+1)T_C.$ For 
$T_C=2.726K$, then $T_{1.776}=7.57K$, which agrees with experiment and
duplicates the standard cosmology's prediction, but without employing the
expansion hypothesis, inflationary or otherwise.

Can the NRI framework also be applied to the interpretation of quasar
redshifts? Perhaps so if we focus on the recent confirmation of the paucity
of quasars with $z>4,$ and their near absence for $z>5.^{14}\mathcal{\ }$
If, for example, we assume $u_g^2\simeq 0.5$ and $v_\theta =0.7c$ might
apply to some of the most distant quasars, then Eq. (5) reveals that $z$
increases from $4.3$ at $8.8 \times 10^9$ ly to $534$ at $9.1263 \times 10^9$
ly. Could such a rapid increase in z be interpreted as one reason for the
sparsity of very high redshift quasars? If so, quasars with $z>4.3$ might
have proper motions of $\sim $ $15$ $\mu as/y$.

Next, consider Hubble Deep Field. It is of interest to note that its
analysis shows that angular diameters of the most distant galaxies do not go
through a minimum and then increase as predicted by the standard cosmology.$%
^{15}$ Instead angular diameters continue to diminish. Standard cosmology's
prediction is of course traceable to its underlying assumption of spacetime
expansion. Since the NRI framework does not have that assumption, it can
quite naturally interpret Hubble Deep Field results as evidence of the
applicability of the general astronomical rule-of-thumb for galaxies---the
further the distance, the smaller the angular diameter.

The NRI framework's apparent ability to interpret a variety of astronomical
and astrophysical observations with Einstein's gravitational and Doppler
redshifts, without the addition of the Friedmann-Lemaitre spacetime
wavelength expansion hypothesis and the Cosmological Principle, was such a
surprise that it seemed quite natural to extend this study and reanalyze the
physics underlying the expansion hypothesis itself.

Our ongoing investigations on this topic have led us to uncover an
intriguing, new test of the expansion hypothesis.$^{16}$ The end result is
yet another surprise almost on par with finding that the NRI framework is an
alternate interpretation of the Hubble relation and the 2.7K CBR.


\begin{thebibliography}{00}
\bibitem{9} Gravity in the NRI is the sum of local and universal
potentials. Thus two
clocks at distances s$_{1{ }}$and s$_2$ from Earth's center should run
at rates that differ by $\sqrt{1+2[\Phi (0)+\varphi (s_1)]/c^2}\ /\sqrt{%
1+2[\Phi (0)+\varphi (s_2)]/c^2}$, where $\varphi (s)$ is the Earth's local
potential. The synchronization of GPS atomic clocks with the$\ \sqrt{%
1+2\varphi (s_1)/c^2]}/\sqrt{1+2\varphi (s_2)/c^2}$ ratio is consistent with 
$\Phi (0)=0.$
\end{thebibliography}
\end{document}